\documentclass[twocolumn,showpacs,amsmath,amssymb]{revtex4}
\usepackage{graphicx}% Include figure files
\usepackage{dcolumn}% Align table columns on decimal point
\usepackage{bm}% bold math
\usepackage{stmaryrd}
\usepackage{wasysym}
\usepackage[figtopcap]{subfigure}

\preprint{APS/123-QED}
\begin{document}

\title{Hyperfine Paschen-Back regime in alkali metal atoms: consistency of two theoretical considerations and experiment}

\author{A. Sargsyan$^1$}
%\email{}
\author{G. Hakhumyan$^1$}
\author{C. Leroy$^2$}
\author{Y. Pashayan-Leroy$^2$}
\author{A. Papoyan$^1$}
\author{D. Sarkisyan$^1$}
\author{M. Auzinsh$^3$}
\affiliation{$^1$Institute for Physical Research, 0203, Ashtarak-2, Armenia}
\affiliation{$^2$Laboratoire Interdisciplinaire Carnot de Bourgogne, UMR CNRS 6303, Universit\'e de Bourgogne, 21078 Dijon Cedex, France}
\affiliation{$^3$Department of Physics, University of Latvia, 19 Rainis Blvd., Riga LV-1586, Latvia}

\date{\today}
\begin{abstract}
Simple and efficient ``$\lambda$-method'' and ``$\lambda/2$-method'' ($\lambda$  is the resonant wavelength of laser radiation) based on nanometric-thickness cell filled with rubidium are implemented to study the splitting of hyperfine transitions of $^{85}$Rb and $^{87}$Rb $D_1$ line in an external magnetic field
in the range of $B = 0.5  - 0.7$~T. It is experimentally demonstrated from
20 (12) Zeeman transitions allowed at low $B$-field in $^{85}$Rb ($^{87}$Rb) spectra
in the case of  $\sigma^+$ polarized laser radiation, only 6 (4) remain at $B > 0.5~\textmd{T}$,
caused by decoupling of the total electronic momentum $J$ and the nuclear spin
momentum $I$ (hyperfine Paschen-Back regime). The expressions derived in the frame
of completely uncoupled basis ($J, m_J ; I, m_I$)  describe very well the experimental
results for $^{85}$Rb transitions at $B > 0.6~\textmd{T}$ (that is a manifestation of hyperfine Paschen-Back regime).
A remarkable result is that the calculations based on the eigenstates of coupled ($F, m_F$) basis,
which adequately describe the system for low magnetic field, also predict reduction of number of
transition components from 20 to 6 for $^{85}$Rb, and from 12 to 4 for $^{87}$Rb spectrum at
$B > 0.5~\textmd{T}$. Also, the Zeeman transitions frequency shift, frequency interval
between the components  and their slope versus $B$ are in agreement with the experiment.

\end{abstract}

\maketitle

\section{Introduction}
Recently it was demonstrated that optical nanometric-thin cell (NTC) containing
atomic vapor of alkali metal (Rb, Cs, etc) allows one to observe a number of spectacular effects,
 which are not observable in ordinary (centimeter-length) cells, particularly:
 1) cooperative effects such as the cooperative Lamb shift caused by dominant
 contribution of atom-atom interactions~\cite{Lamb}; 2) negative group index
 $n_g = -10^5$ (the largest negative group index measured to date) caused by
 propagation of near-resonant light through a gas with $L= \lambda/2$ thickness but many
 atoms per  $\lambda^3$~\cite{SuperLum}; 3) broadening and strong shifts of resonances,
 which become significant when $L \sim 100$~nm caused by atom-surface van der Waals
 interactions due to the tight confinement in NTC~\cite{VdW}.\\
\indent Atomic spectroscopy with NTCs was found to be efficient also for studies
of optical atomic transitions in external magnetic field manifested in two interconnected effects:
splitting of atomic energy levels to Zeeman sublevels (deviating from the linear dependence
in quite moderate magnetic field), and significant change in probability of atomic transitions
as a function of $B$-field~\cite{Cyr,Khv,Auz,Hap,Bud,Hug}. The efficiency of NTCs for
quantitative spectroscopy of Rb atomic levels in magnetic field up to $0.7~\textmd{T}$
has been shown recently~\cite{Sarg2008,Sarg2012}. These studies benefited from
the following features of NTC: 1) sub-Doppler spectral resolution for atomic vapor thickness
$L = \lambda$ and $L = \lambda/2$ ($\lambda$ being the resonant wavelength of Rb~$D_1$
 or $D_2$ line, 795 or 780 nm, respectively) needed to resolve large number of Zeeman
 transition components in transmission or fluorescence spectra; 2) possibility
 to apply strong magnetic field using permanent magnets: in spite of the
 strong inhomogeneity of $B$-field (in our case it can reach $15~\textmd{mT/mm}$),
 the variation of $B$-field inside atomic vapor is negligible because of the small thickness.\\
 \indent Two considerations have been used for theoretical description of behavior
 of the atomic states exposed to strong magnetic field: coupled ($F, m_F$) basis,
 and uncoupled ($J, m_J; I, m_I$) basis, where $J$ is the total electronic angular
 momentum, $I$ is the nuclear spin momentum, $F = I + J$, and $m_J$, $m_I$, and $m_F$, are
 corresponding projections. The completely uncoupled basis is valid for strong
 magnetic field given by $B \gg B_0 = A_{hfs}/\mu_B$, where $A_{hfs}$ is the ground-state
 hyperfine coupling coefficient, $\mu_B$ is the Bohr magneton.
 This regime is called hyperfine Paschen-Back regime (HPB)~\cite{Hap,Steck}.

\section{Theoretical model}
\label{sec:Theory}
If we have an atom with the electronic angular momentum $J$ and nuclear spin $I$,
due to hyperfine interaction between the electronic and nuclear angular momentum,
 atomic fine structure levels are split into the hyperfine components represented by
 the total angular momentum $F$. If an external magnetic field is applied coupling between
 electronic and nuclear angular momentum gradually is destroyed and finally at a very strong
 magnetic field both electronic and nuclear angular momenta interact with the magnetic field
 independently. This means that at a very weak magnetic field the most convenient way to describe
 an atom in a magnetic field is a coupled basis approach which assumes that both angular momenta
 are strongly coupled. This approach is called coupled basis formalism and it uses the basis
 which we will represent in a form

\begin{equation} \label{Eq1}
           |  (J I) F m_{F}  \rangle,
\end{equation}

\noindent where $m_F$ is the magnetic quantum number for hyperfine momentum.

In contrary in a very strong magnetic field when both angular momenta are totally uncoupled
the most convenient is the uncoupled bases approach when the eigenfunctions of an atomic state can be represented as

\begin{equation}
            | J m_{J} \rangle| I m_{I} \rangle ,
\end{equation}

\noindent where $m_J$ and $m_I$ are the magnetic quantum numbers for electronic and nuclear angular momentum respectively.

Of course, both basis according to the quantum angular momentum theory are related via $3jm$ symbols in a simple way \cite{Var}

\begin{align}
            | (J I) F m_{F}   \rangle & =  (-1)^{J-I+m_{F}} \sqrt{2 F +1} \times \nonumber \\
& \times  \sum_{m_J m_I}
\left(
\begin{array} {ccc}
 J & I & F \\
 m_{J} & m_{I} & -m_{F}
\end{array}
\right)
| J m_J \rangle | I m_{I} \rangle ,
\end{align}

\begin{align}
| J m_J \rangle | I m_I \rangle & =
(-1)^{J-I+m_{F}} \sqrt{2 F +1} \times \nonumber \\
& \times \sum_{F m_F}
\left(
\begin{array} {ccc}
 J & I & F \\
 m_{J} & m_{I} & -m_{F}
\end{array}
\right)
| (J I) F m_{F}   \rangle ,
\end{align}
where quantities in brackets are $3jm$ symbols.

If we need to calculate the eigenvalues and eigenfunctions of such an atom in an external
magnetic field of intermediate strength, than of course, neither of the basis
are eigenfunctions of the Hamilton operator which for an atom with the hyperfine interaction can be written as

\begin{equation}
\hat{H} = \hat{H}_{0} + \hat{H}_{hfs} + \hat{H}_{B} ,
\end{equation}

\noindent where $\hat{H}_{0}$ is a Hamilton operator for the unperturbed atom. In our case we are assuming that it is the fine structure state of an atom. The $\hat{H}_{hfs}$ is the hyperfine interaction operator and finally $\hat{H}_{B}$ is the Hamilton operator responsible for the interaction of the atom with an external magnetic field $\textbf{{\textit{B}}}$. Explicitly the hyperfine interaction operator accounting for the magnetic dipole -- dipole interaction and the electric quadrupole interaction between nuclear and electronic angular momenta, can be written as \cite{Steck}

\begin{equation}
\hat{H}_{hfs} = A_{hfs} \hat{I} \hat{J} + B_{hfs} \frac{3 ( \hat{I} \hat{J})^2 + \frac{3}{2} (\hat{I} \hat{J}) - I(I+1)J(J+1)}{2I(2I-1)J(2J-1)} , \label{hfop}
\end{equation}

\noindent where $B_{hfs}$ is an electric quadrupole interaction constant. For simplicity we are neglecting here the higher multiple interaction terms, which usually are much smaller.

The Hamilton operator responsible for the interaction of an atom with the magnetic field can be written as

\begin{equation}
\hat{H}_{B} = - \hat{\mu}_{J} \hat{B} -  \hat{\mu}_{I} \hat{B} = g_{J} \frac{\mu_{B}}{\hbar} \hat{J} \hat{B} +  g_{I} \frac{\mu_{B}}{\hbar} \hat{I} \hat{B} ,
\end{equation}

\noindent where $\hat{\mu}_J$ and $\hat{\mu}_I$ are the magnetic moment operators for electronic and nuclear part of an atom.

If we are interested to find eigenfunctions and energies of atomic levels in the
intermediate strength fields we should calculate these eigenvalues and eigenfunctions
of Hamilton matrix calculated with one of the basis describe above. Each option
has its technical advantages and disadvantages, but both options will give exactly the same result.
Even more, these results can be considered as exact until the additional energy in the external
magnetic field can be considered as small in comparison to the fine structure splitting of atomic states.

If we are using coupled state basis, Hamilton matrix related to the hyperfine interaction will be diagonal,
but magnetic interaction will give the off-diagonal elements. If on contrary we are using uncoupled basis wave functions,
than hyperfine interaction operator will be contributing off-diagonal elements, but magnetic field part will be diagonal.

For example in a coupled basis, diagonal and non diagonal elements responsible for interaction with the magnetic field can be found using the relation~\cite{Auz}

\begin{align}
\langle (JI) F_{i} m_{F} | \bold{J} | (JI) F_{k} m_{F} \rangle  =  (-1)^{J+I+F_{i} + F_{k} - m_{F} + 1}  \times \nonumber \\
 \times \sqrt{(2F_{i} +1)(2F_{k} +1)J(J+1)(2J+1)} \times \nonumber \\
\times  \left(
 \begin{array} {ccc}
 F_{i} & 1 & F_{k} \\
 -m_{F} & 0 & m_{F}
\end{array}
\right)
\left\{
\begin{array} {ccc}
J & F_{i} & I \\
F_{k} & J & 1
\end{array}
\right\}
\end{align}

and

\begin{align}
\langle (JI) F_{i} m_{F} | \bold{I} | (JI) F_{k} m_{F} \rangle  =  (-1)^{J+I+F_{i} + F_{k} -m_{F} + 1}  \times \nonumber \\
 \times \sqrt{(2F_{i} +1)(2F_{k} +1)I(I+1)(2I+1)} \times \nonumber \\
\times  \left(
 \begin{array} {ccc}
 F_{i} & 1 & F_{k} \\
 -m_{F} & 0 & m_{F}
\end{array}
\right)
\left\{
\begin{array} {ccc}
I & F_{i} & J \\
F_{k} & I & 1
\end{array}
\right\} ,
\end{align}

\noindent where quantities in brackets are $3jm$ symbols, and in curled brackets $6j$ symbols.
Hyperfine interaction matrix in this basis is diagonal and its matrix elements are energies of
the hyperfine states. These diagonal matrix elements can be found to be equal to~\cite{Steck}

\begin{equation}
E_{hfs} = \frac{1}{2} A_{hfs} K + B_{hfs} \frac{\frac{3}{2} K(K+1) -2I(I+1)J(J+1)}{4I(2I-1)J(2J-1)} ,
\end{equation}

\noindent where

\begin{equation}
K=F(F+1)-I(I+1)-J(J+1)
\end{equation}

If on the contrary we have decided to start our calculations with the uncoupled bases states,
then the magnetic field operator now is diagonal with matrix elements equal to

\begin{equation}
\label{Eu}
E_{|J,m_J,I,m_I\rangle}=A_{hfs}m_Jm_I+\mu_B(g_Jm_J+g_Im_I)B.
\end{equation}

\noindent where $g_J$ and $g_I$ are Land\'{e} factors for electronic structure of an atom and for nucleus.
The hyperfine interaction matrix is non diagonal for uncoupled basis. To calculate it one must
have the matrix elements for the $\hat{I} \hat{J}$ operator, see Eq. (\ref{hfop}).
Taking into account, that according to the cosine law

\begin{equation}
2(\hat{I}\hat{J}) = \hat{F}^2 - \hat{I}^2 - \hat{J}^2 ,
\end{equation}

\noindent these matrix elements can be found as \cite{Auz}

\begin{align}
\langle J m_J^{'} | \langle I m_I^{'} | \hat{I} \hat{J} | J m_J \rangle | I m_I \rangle  = \nonumber \\
 = \frac{1}{2} \sum_F
(-1)^{2J-2I+m_J +m_J^{'}+m_I +m_I^{'}} (2F+1) \times \nonumber \\
\times \left(
\begin{array} {ccc}
 J & I & F \\
 m_{J} & m_{I} & -m_{J} -m_{I}
\end{array}
\right)
\left(
\begin{array} {ccc}
 J & I & F \\
 m_{J}^{'} & m_{I}^{'} & -m_{J}^{'} -m_{I}^{'}
\end{array}
\right) \times \nonumber \\
\times [F(F+1) - J(J+1) - I(I+1)] .
\end{align}

One must conclude that coupled basis approach is preferable if we
have very weak magnetic field and additional energy that atomic level
gains in the magnetic field is much smaller than the hyperfine energy splitting.
Than we can assume that Zeeman effect is linear and additional magnetic energy can be calculated as

\begin{equation}
\Delta E = g_{F} \mu_B B m_F , \label{Ec}
\end{equation}

\noindent where $g_F$ is the hyperfine Land\'{e} factor~\cite{Steck}.
%
%\begin{align}
%g_F = g_J \frac{F(F+1) + J(J+1) - I(I+1)}{2F(F+1)} + \nonumber \\
%+ g_I \frac{F(F+1) + I(I+1) - J(J+1)}{2F(F+1)} .
%\end{align}

On contrary the uncoupled basis is preferred when the magnetic
field is large enough $B \gg B_0$ to assume that the electronic
and nuclear angular momentum are uncoupled. Then the additional
energy of an atom in the magnetic field can be simply calculated according to the equation (\ref{Eu}).

\section{Experimental}
\subsection{Nanometric-thin cell}
\begin{figure}[hbtp]
% \centering
 \includegraphics[scale=0.5] {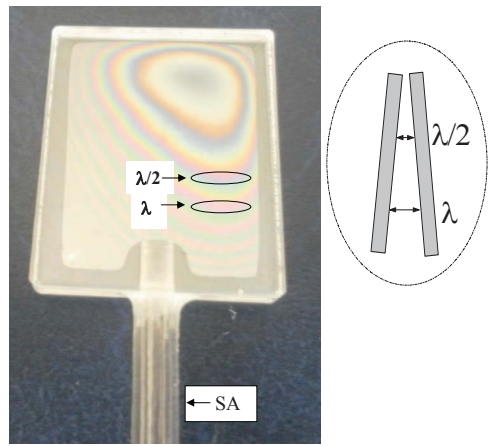}
 \caption{
 Photograph of the nanometric-thin cell with vertically wedged vapor gap.
 Regions of $L = \lambda/2 = 397.5$~nm and $L = \lambda = 795$~nm are marked.
  SA is the sapphire side-arm filled with metallic Rb.
 }
 \label{fig:NTC}
 \end{figure}

NTCs filled with Rb have been used in our experiment, which allowed to
obtain sub-Doppler spectra and resolve hyperfine and Zeeman atomic components.
The general design of nanometric-thin cell was similar to that of extremely
thin cell described earlier~\cite{Sark2001,Sarg2011}. The rectangular $20 \times 30$~mm$^2$, 2.5~mm-thick window
wafers polished to less than $1$~nm surface roughness were fabricated from commercial
sapphire (Al$_2$O$_3$), which is chemically resistant to hot vapors
(up to $1000$~$^{\circ}$C) of alkali metals.
The wafers were cut across the c-axis to minimize the birefringence.
In order to exploit variable vapor column thickness, the cell was vertically
wedged by placing a $1.5$~$\mu$m-thick platinum spacer strip between the windows
at the bottom side prior to gluing. The NTC is filled with a natural
rubidium ($72.2\%$~$^{85}$Rb and $27.8\%$~$^{87}$Rb).
The photograph of the NTC cell is presented in Fig.~\ref{fig:NTC}.
 Since the gap thickness $L$ between the inner surfaces of the windows
 (the thickness $L$ of Rb atomic vapor column) is of the order of visible light wavelength,
 one can clearly see an interference pattern visualizing smooth thickness variation
 from $50$ nm to $1500$ nm. The NTC behaves as a low finesse Fabry-P\'{e}rot etalon, and
 the reflection $R$ of the NTC can be described by formulas for the thickness dependence
 of reflected power. The latter has been exploited for the precise measurement of
 the vapor gap thickness across the cell aperture. Particularly, $R \approx  0$
 when $L = n\lambda/2$ ($n$ is integer), which is very convenient for the
 experimental adjustment. The accuracy of the cell thickness measurement is better than $20$ nm.

 \noindent The NTC operated with a special oven with four optical outlets: a pair of in
 line ports for laser beam transmission and two orthogonal ports to collect the side
 fluorescence. This geometry allows simultaneous detection of transmission and fluorescence spectra.
 The oven with the NTC fixed inside was rigidly attached to a translation stage for
 smooth vertical translation to adjust the needed vapor column thickness without
 variation of thermal conditions. A thermocouple is attached to the sapphire side
 arm (SA) at the boundary of metallic Rb to measure the temperature, which determines
 the vapor pressure. The SA temperature in present experiment was set to $120$~$^{\circ}$C,
 while the windows temperature was kept some $20$~$^{\circ}$C higher to prevent condensation.
 This regime corresponds to Rb atomic density $N = 2\times 10^{13}$ cm$^{-3}$.
\subsection{Experimental arrangement}
\begin{figure}[hbtp]
% \centering
 \includegraphics[scale=0.3] {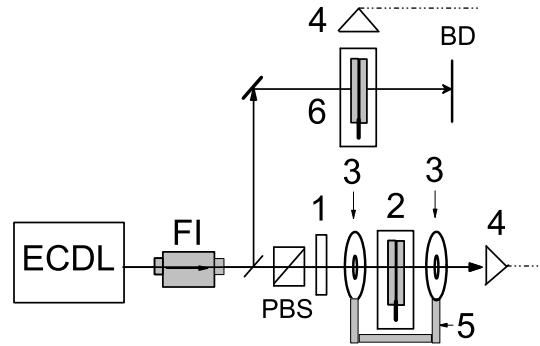}
 \caption{
 Sketch of the experimental setup. ECDL: diode laser; FI: Faraday isolator;
 1: $\lambda/4$ plate; 2: NTC in the oven; PBS: polarizing beam splitter;
 3: permanent ring magnets; 4: photodetectors; 5: stainless steel $\Pi$-shape holder;
 6: auxiliary Rb NTC with thickness $L = \lambda/2$; BD-beam dumper.
 }
 \label{fig:SetUp}
 \end{figure}

Sketch of the experimental setup is presented in Fig.~\ref{fig:SetUp}.
The circularly polarized beam of extended cavity diode
laser ($\gamma_L < 1$~MHz) resonant with Rb $D_1$ line, after passing
through Faraday isolator was focused to a $0.3$~mm diameter spot
onto the Rb NTC (2) orthogonally to the cell window.  A polarizing beam
splitter (PBS) was used to purify initial linear polarization of the laser
 radiation; a $\lambda/4$ plate (1) was utilized to produce a circular polarization.
 In the experiments the thicknesses of vapor column $L = \lambda$ and $L = \lambda/2$
 have been exploited. The transmission and fluorescence spectra were
 recorded by photodiodes with amplifiers followed by a four channel
 digital storage oscilloscope, Tektronix TDS 2014B. To record
 transmission and fluorescence spectra, the laser radiation was
 linearly scanned within up to $20$~GHz spectral region covering the
 studied group of transitions. The linearity of the scanned frequency
 was monitored by simultaneously recorded transmission spectra of a
 Fabry-P\'{e}rot etalon (not shown). The nonlinearity has been evaluated
 to be about 1$\%$ throughout the spectral range. About 30$\%$ of the pump
 power was branched to the reference unit with an auxiliary Rb NTC (6).
 The fluorescence spectrum of the latter with thickness $L = \lambda/2$ was
 used as a frequency reference for $B = 0$~\cite{Sark2004}.\\
 \indent The assembly of oven with NTC inside with $8$~mm longitudinal size
 was placed between the permanent ring magnets. Magnetic field was
 directed along the laser radiation propagation direction $\textbf{k}$
 ($\textbf{B} \sslash \textbf{k}$).
 Extremely small thickness of the NTC is advantageous for the application
 of very strong magnetic fields with the use of permanent magnets having
 a $2$~mm diameter hole for laser beam passage. Such magnets are unusable for
 ordinary cm-size cells because of strong inhomogeneity of the magnetic
 field, while in NTC, the variation of the $B$-field inside the atomic
 vapor column is several orders less than the applied $B$ value.
 The permanent magnets are mounted on a $\Pi$-shaped holder with $50\times50$~mm$^2$
 cross-section made from soft stainless steel. Additional form-wounded
 copper coils allow the application of extra $B$ field (up to   $\pm0.1$~T).
 The $B$-field strength was measured by a calibrated Hall gauge
 with an absolute imprecision less than $5$~mT  throughout the applied $B$-field range.

\subsection{Realization of the sub-Doppler resolution: ``$\lambda$-method'' and ``$\lambda/2$-method''}
\label{SubSec:LambdaMethods}
Two different methods based on the NTC were implemented to study the behavior
of frequency-resolved individual atomic Zeeman transitions exposed to external magnetic field.\\
\indent
1)``$\lambda$-method". As it was shown in~\cite{Sark2004,Duc}, the NTC with thickness of Rb
atomic vapor column $L =\lambda$, with $\lambda = 795$~nm being the wavelength of the
laser radiation resonant with the Rb $D_1$ line, is an efficient tool to
attain sub-Doppler spectral resolution. Spectrally narrow (10-15~MHz)
velocity selective optical pumping (VSOP) resonances located exactly at
the positions of atomic transitions appear in the transmission spectrum
of NTC at the laser intensities   $10$ mW/cm$^2$. The VSOP parameters are shown
to be immune against 10$\%$ thickness deviation from $L = \lambda$, which makes
 ``$\lambda$-method"
feasible. When NTC is placed in a weak magnetic field, the VSOPs are split
into several components depending on ($F, m_F$), while in the case of strong
magnetic fields the VSOPs numbers are determined by the ($J, m_J; I, m_I$)
quantum numbers.
The amplitudes and frequency positions of VSOPs depend on the $B$-field,
which makes it convenient to study separately each individual atomic transition~\cite{Sarg2008}.\\
\indent 2)``$\lambda$/2-method". This technique exploits strong narrowing in absorption spectrum
at $L = \lambda/2$ as compared with the case of an ordinary cm-size cell [18]. Particularly,
the absorption linewidth for Rb $D_1$ line reduces to $\sim$~120~MHz FWHM (Full Width Half Maximum), as
opposed to $\sim$~500~MHz in an ordinary cell. The absorption profile in the
case of $L = \lambda/2$ is described by a convolution of Lorentzian and Gaussian
profiles (Voigt profile). The sharp (nearly Gaussian) absorption near the
top makes it convenient to separate closely spaced individual atomic
transitions in an external magnetic field. Also in this case the deviation
of thickness by 10$\%$ from $L = \lambda/2$ weakly effects the absorption linewidth.
We have used advantages of ``$\lambda$-method" and ``$\lambda/2$-method" throughout our studies presented below.

\section{Consistency of experiment with theoretical considerations}
\subsection{Studies for $^{85}$Rb and $^{87}$Rb by ``$\lambda$-method": $B = 0.5  - 0.7$~T}
\begin{figure}[hbtp]
% \centering
 \includegraphics[scale=0.3] {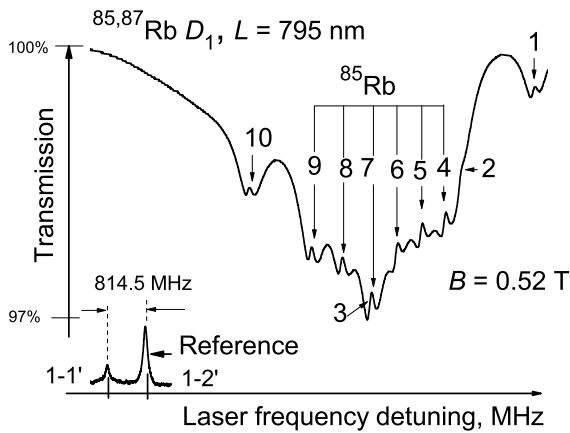}
 \caption{
 Transmission spectrum of Rb NTC with $L = \lambda$ for $B = 0.52$~T and $\sigma^{+}$ laser excitation.
 The VSOP resonances marked $4-9$ belong to $^{85}$Rb; resonances marked $1, 2, 3, 10$
 belong to $^{87}$Rb. The lower curve is fluorescence spectrum of the reference
 NTC with $L = \lambda/2$, showing the positions of $^{87}$Rb $F_g=1 \rightarrow F_e=1,2$ transitions
 for $B = 0$, labeled as $1-1^{\prime}$ and $1-2^{\prime}$.
 }
 \label{fig:Fig3}
 \end{figure}

The estimates for a $B$-field required to decouple the total electronic
angular momentum and the nuclear spin momentum defined by $B \gg B_0= A_{hfs}/\mu_B$
give $B_0= 0.07$~T for $^{85}$Rb and $B_0=0.2$~T for $^{87}$Rb. The recorded transmission
spectrum of Rb NTC with thickness $L = \lambda$ for $\sigma^{+}$ laser excitation
and $B = 0.52$~T is shown in Fig.~\ref{fig:Fig3}.
The VSOP resonances labeled $1 -10$ demonstrate increased transmission at
the positions of the individual Zeeman transitions: six transitions,
$4-9$, belong to $^{85}$Rb, and four transitions, $1, 2, 3, 10$ belong to $^{87}$Rb.
VSOPs labeled $3$ and $7$ are overlapped. The larger amplitudes for $^{85}$Rb
components are caused by isotopic abundance in natural Rb (72$\%$ $^{85}$Rb,
28$\%$ $^{87}$Rb).
The lower curve shows the fluorescence spectrum of the reference NTC with $L = \lambda/2$,
showing the positions of $^{87}$Rb, $F_g=1 \rightarrow F_e=1, 2$ transitions.
Frequency shifts of all
the VSOP peaks are measured from $F_g=1 \rightarrow F_e=2$ transition.
\begin{figure}[hbtp]
% \centering
 \includegraphics[scale=0.3] {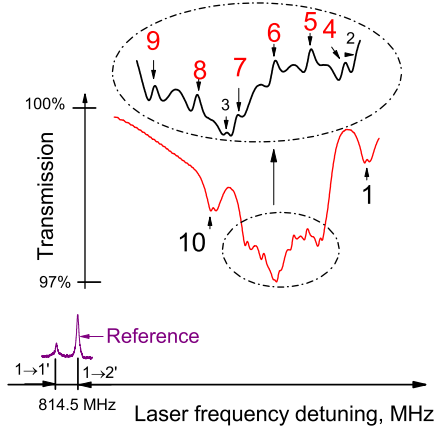}
 \caption{
 Transmission spectrum of Rb NTC with $L = \lambda$ for $B = 0.677$~T and $\sigma^{+}$ laser excitation.
 The labeling of VSOP resonances is the same as in Fig.~\ref{fig:Fig3}. All the VSOP
 resonances are well resolved. The lower curve is the fluorescence spectrum of the reference
 NTC with $L = \lambda/2$, showing the positions of $^{87}$Rb $F_g=1 \rightarrow F_e=1, 2$
 transitions for $B = 0$.
 }
 \label{fig:Fig4}
 \end{figure}
The further increase of a $B$-field results in complete resolving of all the
transition components (including 3 and 7). The transmission spectrum recorded
for $B = 0.677$~T, otherwise in the same conditions as in Fig.~\ref{fig:Fig3} is
presented in Fig.~\ref{fig:Fig4}.\\
\indent As it is mentioned above, in the case of HPB regime the eigenstates of
the Hamiltonian are described in the uncoupled basis of $J$ and $I$ projections ($m_J; m_I$).
Fig.~\ref{fig:Fig5}\subref{fig:Fig5a} presents a diagram of six Zeeman transitions of $^{85}$Rb for the HPB regime in
the case of $\sigma^{+}$ polarized laser excitation (selection rules: $\Delta m_J = +1; \Delta m_I = 0$),
with the same labeling as in Figs.~\ref{fig:Fig3},\ref{fig:Fig4}.
\begin{figure}[t!]
\subfigure[]{
\resizebox{0.45\textwidth}{!}{\includegraphics{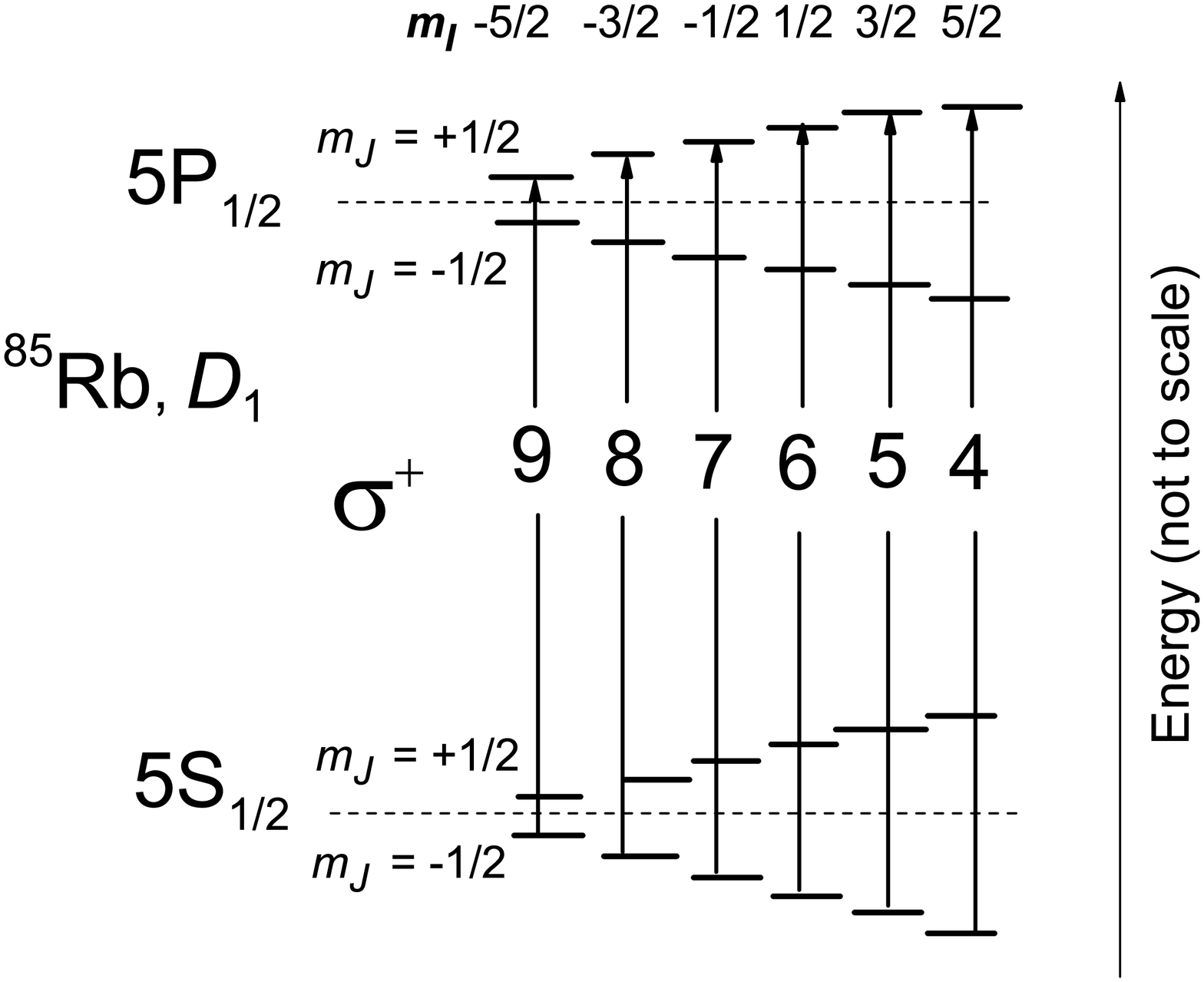}
}
\label{fig:Fig5a}
} \hspace{0.005cm}
\subfigure[]{
\resizebox{0.45\textwidth}{!}{\includegraphics{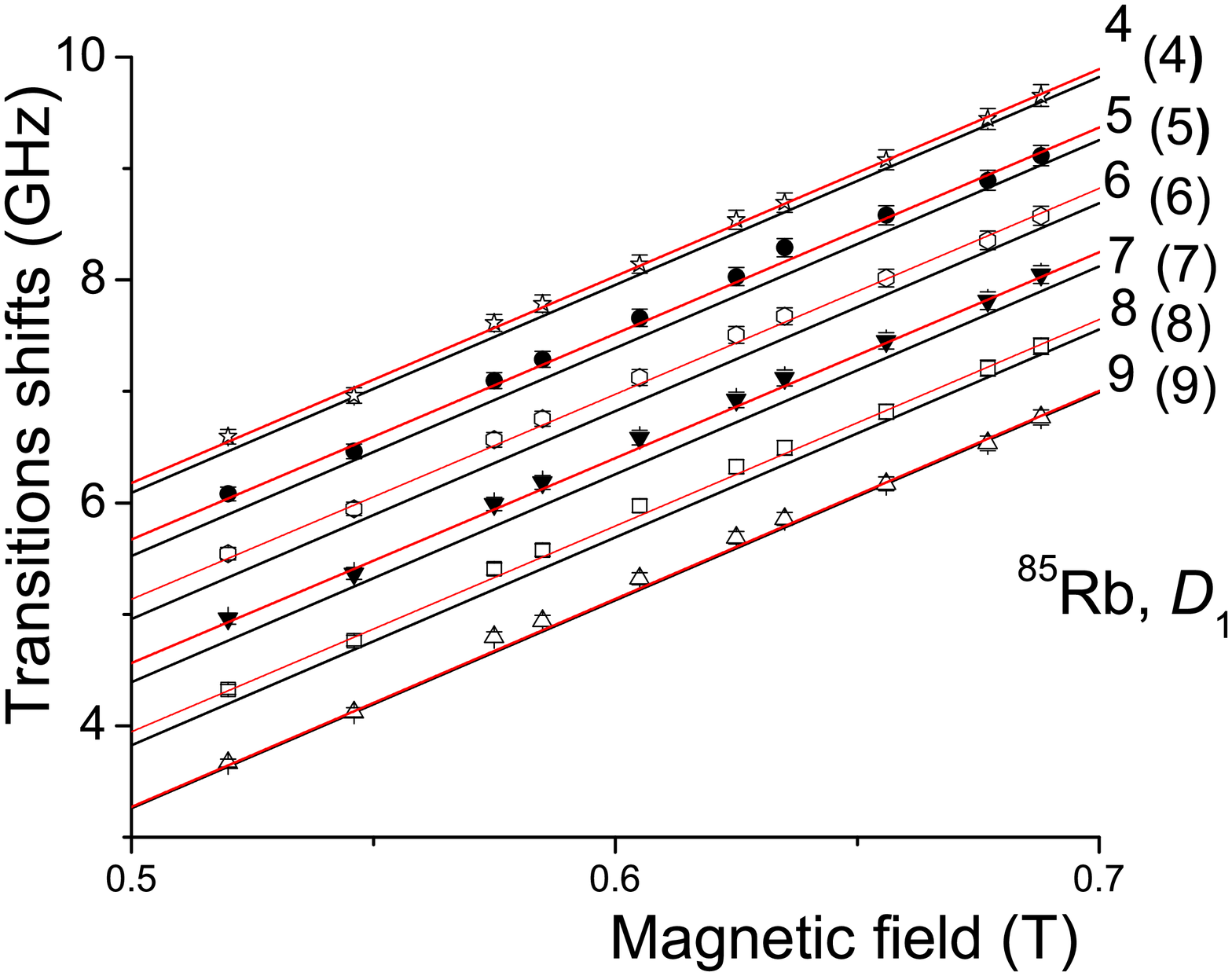}
}
\label{fig:Fig5b}
}
\caption{(Color online) a) Diagram of $^{85}$Rb $D_1$ line transitions in the HPB
regime for $\sigma^{+}$ laser excitation. b) Magnetic field dependence of the frequency
shifts for the transition components $4-9$. Red solid lines $4-9$: calculation by
the coupled basis theory; black solid lines $(4)- (9)$: calculation by the HPB
theory; symbols: experimental results (measurement inaccuracy is   $\pm 1 \%$).
Note, that the curves 9 and (9) are completely overlapped.
 }
\label{fig:Fig5}
\end{figure}
Magnetic field dependence of frequency shift for $^{85}$Rb components $4-9$ is shown in Fig.~\ref{fig:Fig5}\subref{fig:Fig5a}. Red
lines marked $4-9$ are calculated by the coupled basis theory, and black lines $(4)-(9)$ are
calculated by the HPB theory, see Eq.~\ref{Eu}. Symbols represent the experimental results.
As it is seen, for $B > 0.6$~T also the theoretical curves for HPB regime
well describe the experiment with inaccuracy of  $\pm 1\%$.
\begin{figure}[hbtp]
% \centering
 \includegraphics[scale=0.3] {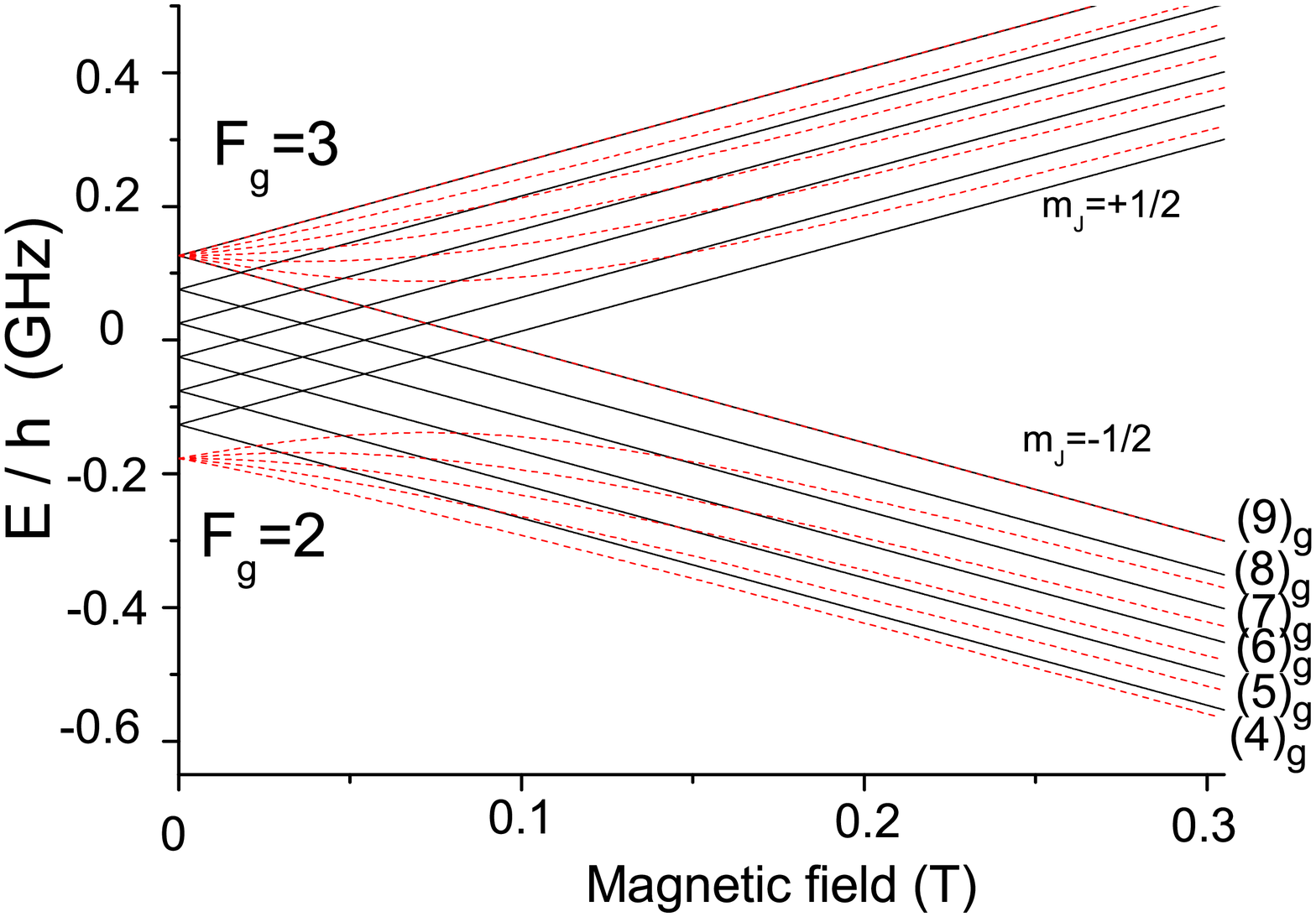}
 \caption{
 (Color online) Theoretical magnetic field dependence of $F_g=2, 3$
 ground hyperfine levels of $^{85}$Rb. Red lines: calculations by the
 coupled basis theory; black lines: calculations as given by Eq.(12) (HPB regime).
 Ground levels for the transitions $4-9$ are indicated as $(4)_g - (9)_g$.
 }
 \label{fig:Fig6}
 \end{figure}
\indent Theoretical graphs for splitting of ground state hyperfine levels $F_g = 2, 3$
of $^{85}$Rb versus magnetic field starting from $B = 0$ calculated by coupled and
uncoupled basis theories are shown in Fig.~\ref{fig:Fig6}. Ground sublevels for
transitions $4-9$ are indicated as $(4)_{g}-(9)_{g}$. A drastic difference between the two
models observed at low magnetic field due to the complete neglecting
of the $J-I$ coupling in Eq.(11) gradually reduces with the increase of
the $B$-field. Five sublevels of $F_g=2$ and seven sublevels of $F_g=3$ in
coupled basis model (red lines) tend to converge to sublevels of two
six-component groups for uncoupled basis model (black lines) with the
increase of magnetic field. For $B \geq0.6$~T, both models become consistent
with the experimental results to an accuracy of   $\pm 1 \%$ (Fig.~\ref{fig:Fig5}\subref{fig:Fig5b}).
It is important to note, that for the upper states of transitions $4-9$,
the convergence of the two models occurs at much lower magnetic field
($B > 0.2$~T), because the hyperfine coupling coefficient $A_{hfs}$
 for 5P$_{1/2}$ of $^{85}$Rb is  $h  \times 120$~MHz, $8$ times smaller than $A_{hfs}$
for 5S$_{1/2}$.\\
\indent Thus, for $B  \geq 0.6$~T simple equation~(\ref{Eu}) could be used for the determination
of the following important parameters of $^{85}$Rb atoms: 1) Frequency positions of atomic
transition components and frequency separation  $\Delta_{nk}$ of $n$-th and $k$-th atomic transition components:
\begin{align}
   \label{FreqPos}
      \Delta_{nk}  = &  \big \{A_{hfs}\left(P_{1/2}\right)m_{J} \left [m_I(n)-m_I(k) \right ]+ \nonumber  \\
       \quad & A_{hfs}\left(S_{1/2}\right)m_{J}\left [m_I(n)-m_I(k) \right ] \big \}
\end{align}
Particularly, the frequency distance between $n=4$ and $k=5$ components is   $566$~MHz,
which coincides with the experimental results at $B > 0.6$~T to $2\%$ accuracy.
2) The slope $S$ in dependence of atomic transition components frequency
on magnetic field, which is the same for all the $6$ components $4-9$,
and can be calculated by the expression
\begin{align}
   \label{Slope}
      S  = & \left [g_J\left (P_{1/2}\right)m_{J} + g_J\left (S_{1/2}\right)m_{J}\right ]\mu_B/B \nonumber  \\
       \quad & \approx 18.6~\textmd{\textmd{MHz}}/\textmd{mT}
\end{align}
(as $g_I \ll g_J$ , we ignore $g_I m_I$ contribution), which coincides well with the experiment.\\
\indent In Fig.~\ref{fig:Fig7}\subref{fig:Fig7a} four transitions of $^{87}$Rb labeled $1-3, 10$ are shown for the
case of $\sigma^{+}$ polarized laser excitation for the HPB regime (selection rules:
$\Delta m_J = +1; \Delta m_I = 0$). The magnetic field dependence of frequency
shift for these components is presented in Fig.~\ref{fig:Fig7}\subref{fig:Fig7b}. The red curves $1-3, 10$ are calculated
by the coupled basis theory, and the black lines $(1)-(3)$ and $(10)$ are calculated
by the HPB theory, Eq.(12). Symbols represent the experimental results.
\begin{figure}[t!]
\subfigure[]{
\resizebox{0.45\textwidth}{!}{\includegraphics{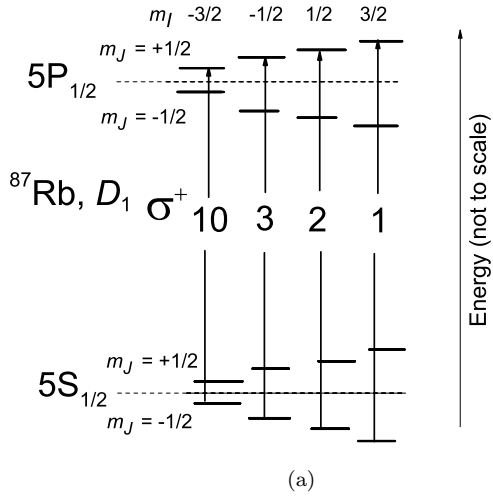}
}
\label{fig:Fig7a}
} \hspace{0.005cm}
\subfigure[]{
\resizebox{0.45\textwidth}{!}{\includegraphics{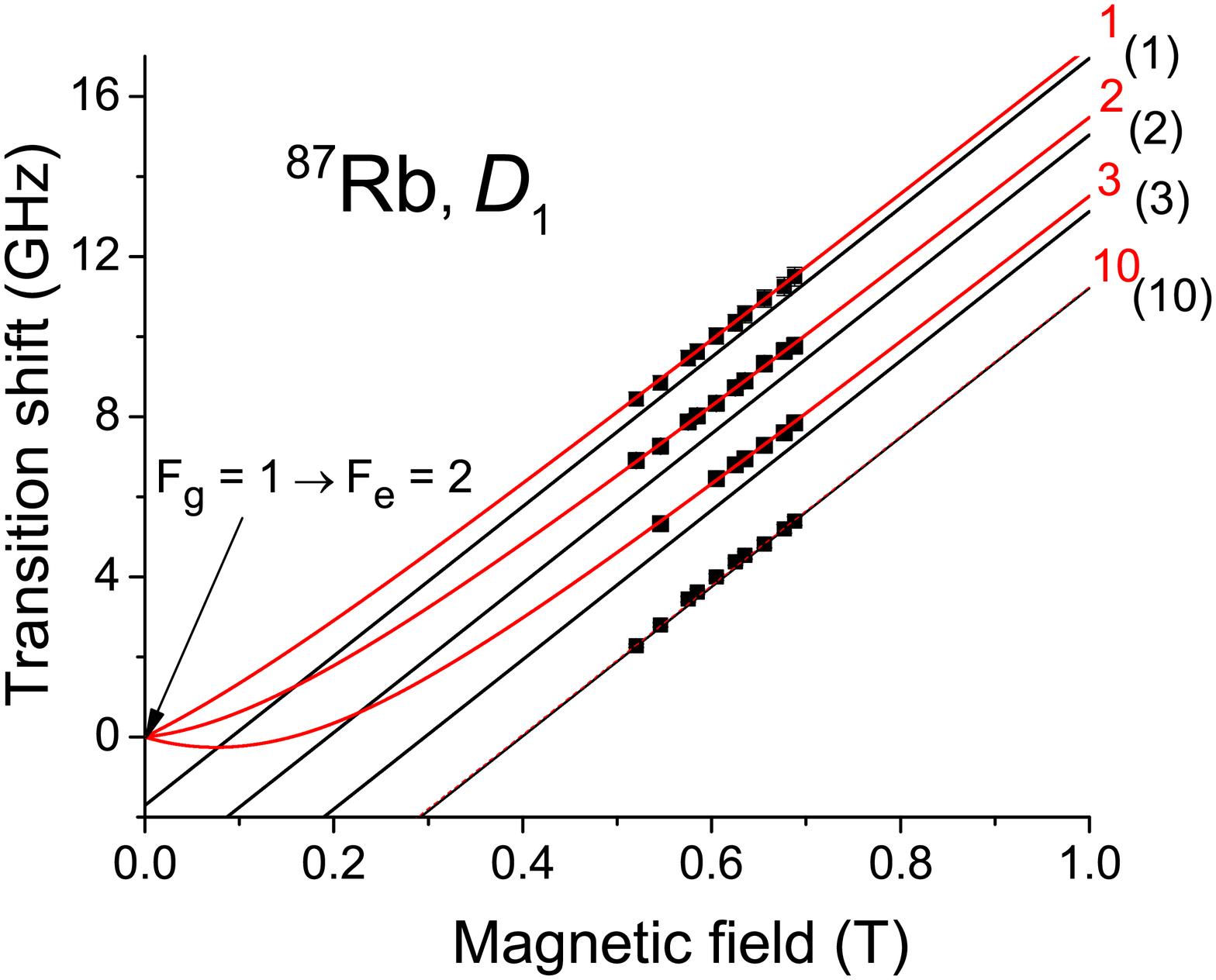}
}
\label{fig:Fig7b}
}
\caption{(Color online) a) Diagram of $^{87}$Rb $D_1$ line transitions in the HPB
regime for $\sigma^{+}$ laser excitation. b) Magnetic field dependence of the frequency
shifts for the transition components $1-3$ and $10$. Red solid lines $1-3, 10$: calculation by
the coupled basis theory; black solid lines $(1)- (3), (10)$: calculation by the HPB
theory; symbols: experimental results (measurement inaccuracy is   $\pm 1 \%$).
Note, that the red curve $10$ and black curve $(10)$ are completely overlapped. }
\label{fig:Fig7}
\end{figure}
Similar to Fig.~\ref{fig:Fig6} and for the same reason, drastic difference between the
two models is observed in Fig.~\ref{fig:Fig7}\subref{fig:Fig7b} for weak magnetic field, with tendency
to converge as the $B$-field increases. However, the curves converge at
significantly higher magnetic field ($> 0.6$~T) required to decouple the nuclear and
electronic spins for $^{87}$Rb having larger hyperfine splitting. It is important to note
that also for four transitions of $^{87}$Rb, the slope $S$ is nearly the same as
for $^{85}$Rb ($S \approx  18.6~\textmd{MHz/mT}$). This is explained by the fact that the expression
for $S$ contains values of $g_J(5S_{1/2})m_J$ and $g_J(5P_{1/2})m_J$ which are the same
for $^{85}$Rb and $^{87}$Rb, but does not contain $A_{hfs}$ values for
$5S_{1/2}$ state that are strongly different.\\
\indent It is worth noting that the complete HPB regime for Cs~$D_2$ line
having the same ground state $A_{hfs}$ value as for $^{87}$Rb, has been observed
in~\cite{Hap} at $B \sim2.7$~T. Thus, one may expect that also for $^{87}$Rb the
 complete HPB regime appears for $B >10 B_0$.
\subsection{Studies of hyperfine Paschen-Back regime for $^{85}$Rb and $^{87}$Rb by ``$\lambda/2$-method"}
\begin{figure}[hbtp]
% \centering
 \includegraphics[scale=0.25] {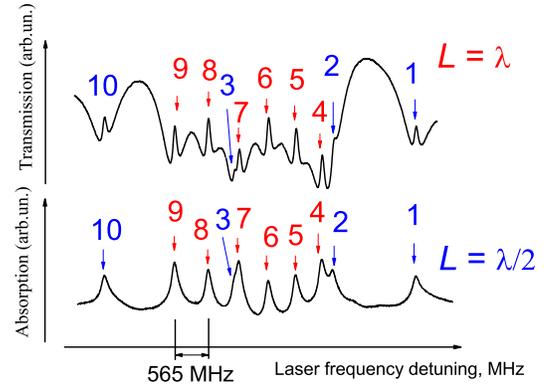}
 \caption{
 (Color online) Comparison of spectra obtained by  ``$\lambda$-method" (upper graph) and ``$\lambda/2$-method"
 (lower graph) for $B = 0.605$~T.
 }
 \label{fig:Fig8}
 \end{figure}

Advantages of  ``$\lambda/2$-method" addressed in Section~\ref{SubSec:LambdaMethods} make it convenient to
separate closely spaced individual atomic transitions in an external magnetic field.
In order to compare  ``$\lambda/2$-method" and   ``$\lambda$-method'' (based on VSOP resonance), we have
combined in Fig.~\ref{fig:Fig8} the spectra obtained by these methods at $B = 0.605~\textmd{T}$, keeping
the previous labeling of individual transitions of $^{85}$Rb and $^{87}$Rb. Let us discuss
 the distinctions of ``$\lambda/2$-method" versus   ``$\lambda$-method". First, it
 requires $4$ orders
 less laser radiation intensity. In the case of low absorption (a few percent),
 the absorption $A$ is proportional to $\sigma NL$, where $\sigma$ is the absorption cross-section
 and is proportional to $d^2$ ($d$ being the dipole moment matrix element),
 $N$ is the atomic density, and $L$ is the thickness. Thus, directly comparing
 $A_i$ (peak amplitudes of the absorption of the  $i$-th transition),
  it is straightforward to estimate the relative probabilities
 (line intensities). Meanwhile for VSOP-based   method the linearity of the response
 has to be verified. Moreover, spatial resolution is twice better for $L = \lambda/2$ as
 compared with $L = \lambda$, which can be important when strongly inhomogeneous magnetic
  field is applied. On the other hand,   method based on VSOP provides 5-fold
  better spectral resolution. Thus, the two methods can be considered as
  complementary depending on particular requirements. Note that it is easy
  to switch from $\lambda/2$ to $\lambda$ in experiment just by vertical translation of the NTC.
\subsection{Consistency of coupled basis model with experiment: $^{85}$Rb }
\begin{figure}[hbtp]
% \centering
 \includegraphics[scale=0.3] {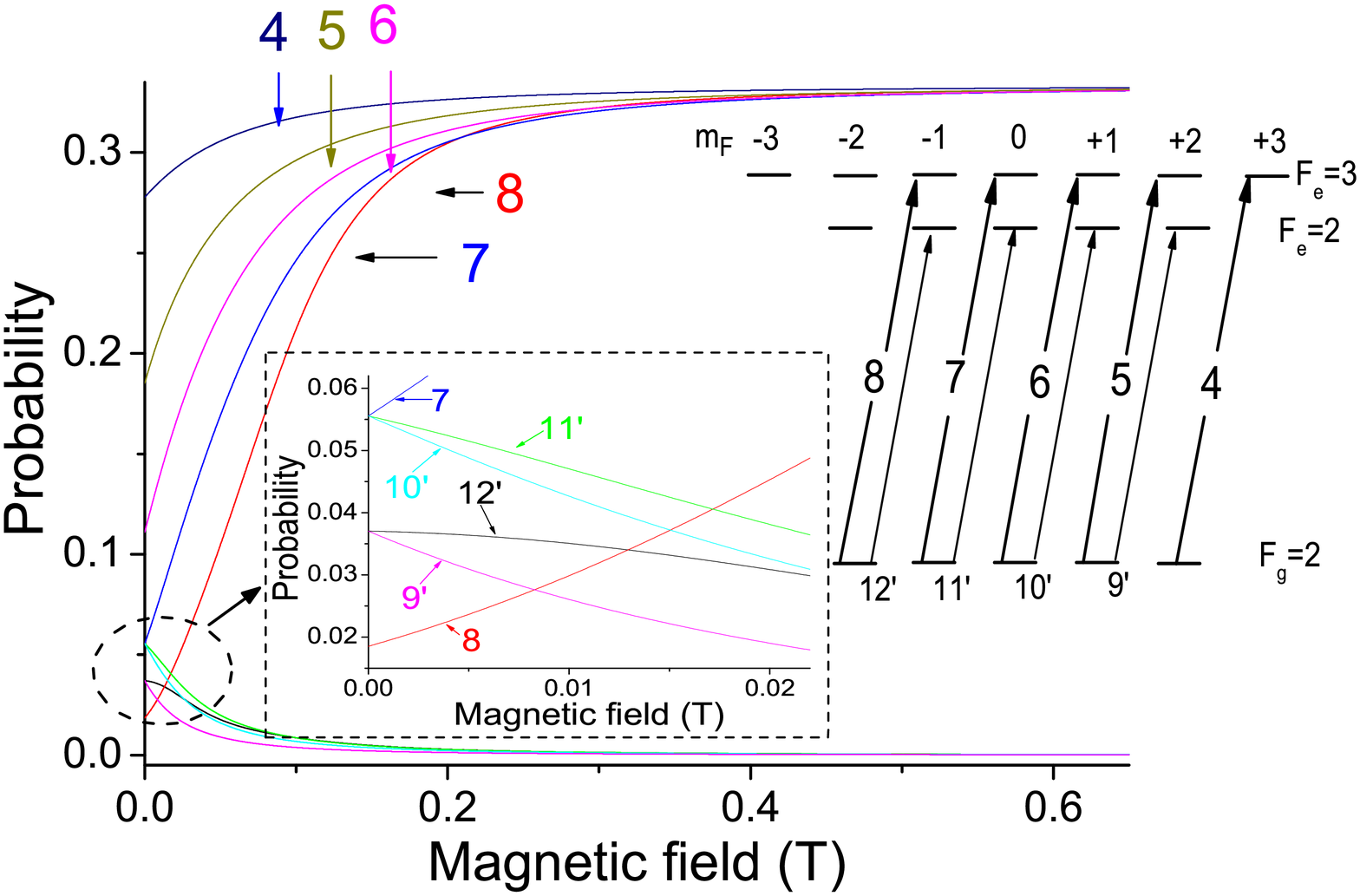}
 \caption{
 (Color online) The probabilities of nine Zeeman components of $F_g=2 \rightarrow  F_e=2, 3$ transitions
 of $^{85}$Rb~$D_1$
 line labeled in the inset versus $B$ for the case of  $\sigma^+ $ excitation.
 }
 \label{fig:Fig9}
 \end{figure}

\begin{figure}[hbtp]
% \centering
 \includegraphics[scale=0.3] {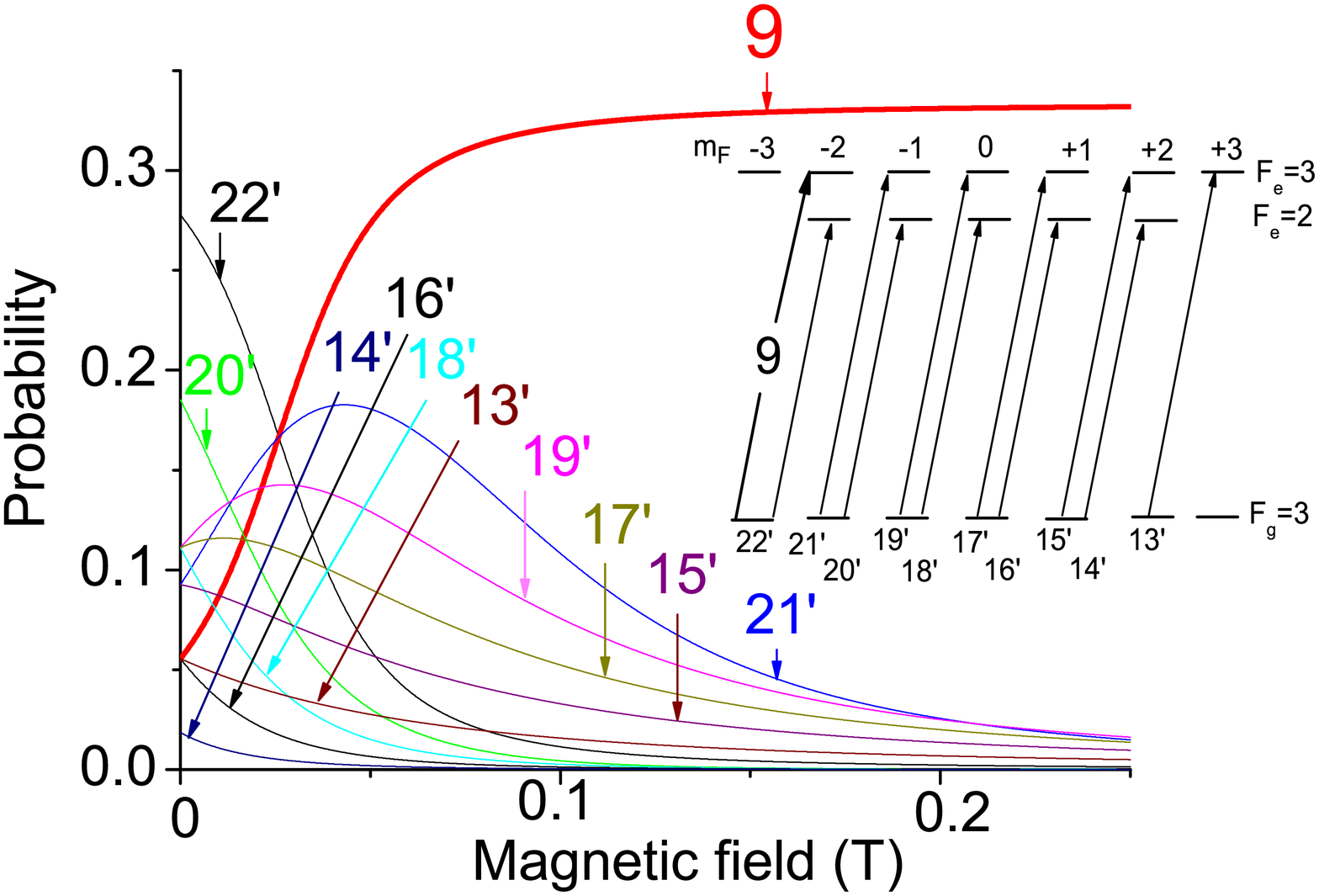}
 \caption{
 (Color online) The probabilities of nine Zeeman components of $F_g=3 \rightarrow  F_e=2, 3$ transitions
 of $^{85}$Rb~$D_1$
 line labeled in the inset versus $B$ for the case of  $\sigma^+ $ excitation.
 }
 \label{fig:Fig10}
 \end{figure}

In the frame of coupled basis for $\sigma^+$ laser excitation, there are twenty
atomic transitions for $^{85}$Rb according to the selection rules. It
should be noted that for $ B < 20~\textmd{mT}$ and  $\sigma^+$ excitation all the twenty
 atomic transitions of $^{85}$Rb have been recorded in~\cite{Auz}.
Fig.~\ref{fig:Fig9} shows the transition probabilities versus $B$ for nine $F_g=2  \rightarrow F_e=2, 3$
transition components under  $\sigma^+$ excitation (see the labeled diagram in the inset).
We can see from Fig.~\ref{fig:Fig9} that the probabilities of transitions $4- 8$ increase, and
probabilities of transitions $9^{\prime}-12^{\prime}$ decrease with $B$, and
for $B > 0.5~\textmd{T}$ only $5$
transitions $(4 -8)$ remain in the spectrum.
Similarly, the probabilities of eleven components of $F_g=3 \rightarrow  F_e=2, 3$ transitions
versus $B$ for the case of  $\sigma^+ $ excitation are presented in Fig.~\ref{fig:Fig10}. Here only the
probability of the transition labeled 9 increases with $B$, remaining the only
component in the spectrum for $B > 0.5$~T.
 Thus, also in the frame of the coupled basis six transitions remain in $^{85}$Rb~$D_1$
 line spectrum at $B > 0.5~\textmd{T}$ for  $\sigma^+ $ excitation.\\
 \indent Although the experimental results obtained for strong magnetic
 field are found to be in consistency with an uncoupled basis model
 (HPB regime) and can be described by simple theoretical expressions
 as is shown in Section~\ref{sec:Theory}, however there are some cases when the coupled
 basis is to be used. Particularly, it was revealed in~\cite{Sarg2008} that
 $F_g=1 \rightarrow  F_e=3$ transition ``forbidden" at $B = 0$ due to the selection rule $\Delta F = 0, \pm1$
 appears in the transmission spectrum of $^{87}$Rb~$D_2$ line at strong magnetic field.
 Even for $B > 0.6~\textmd{T}$, the probability of this transition calculated in the
 coupled basis is not negligible and can be easily detected.
\subsection{Consistency of coupled basis model with experiment: $^{87}$Rb}
\begin{figure}[h]
 \includegraphics[scale=0.25] {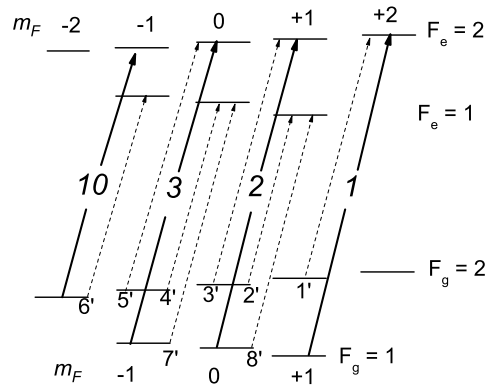}
 \caption{Diagram of $^{87}$Rb~$D_1$ line transitions
 in the frame of coupled basis for $\sigma^+$ laser excitation; the selection rules:
 $\Delta F = 0, 1; \Delta m_F = +1$.
 }
 \label{fig:Fig11}
 \end{figure}

\begin{figure}[h]
 \includegraphics[scale=0.25] {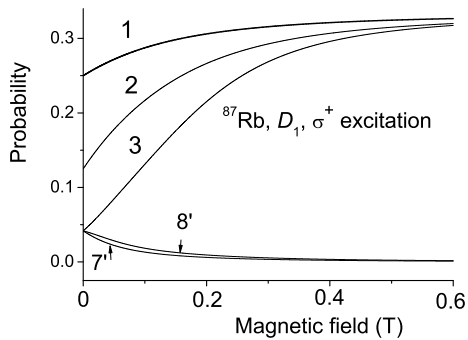}
 \caption{ Calculated probabilities of Zeeman transitions $1-3$, $7^{\prime}$ and $8^{\prime}$
 for  $\sigma^+$ laser excitation versus magnetic field.
 }
 \label{fig:Fig12}
 \end{figure}

\begin{figure}[h]
 \includegraphics[scale=0.25] {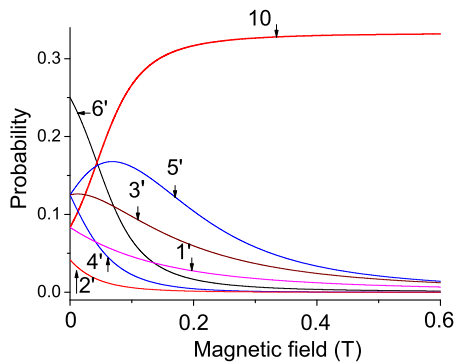}
 \caption{ (Color online) Calculated probabilities of Zeeman transitions $1^{\prime}-6^{\prime}$ and $10$
 for  $\sigma^+$ laser excitation versus magnetic field.
 }
 \label{fig:Fig13}
 \end{figure}

Four atomic transitions of $^{87}$Rb in HPB regime were presented in Fig.~\ref{fig:Fig7}\subref{fig:Fig7a}.
In the frame of coupled basis $(F, m_F)$ for $\sigma^+$ laser excitation there are
twelve atomic transitions according to the selection rules, which are
presented in Fig.~\ref{fig:Fig11}. The transitions labeled $1-3$ and $10$
(shown also in Fig.~\ref{fig:Fig7}\subref{fig:Fig7a})
are depicted by solid lines, and other transitions absent for HPB case are
presented by dashed lines. Note that for weak magnetic field ($B < 20~\textmd{mT}$)
in the case of  $\sigma^+$ excitation all twelve atomic transitions of $^{87}$Rb have
been detected in~\cite{Khv}. In order to find out which atomic transitions will
remain in a strong magnetic field regime, it is needed to calculate the
magnetic field-dependent probabilities for all the twelve atomic transitions.
Fig.~\ref{fig:Fig12} shows the dependence of the probabilities of atomic transitions $1- 5$ on magnetic
field for  $\sigma^+$ laser excitation. It is clearly seen that only transitions $1-3$ remain in
the spectrum for $B > 0.2$~T.
The same dependence for transitions labeled $1^{\prime}-6^{\prime}$ and $10$ is shown in
Fig.~\ref{fig:Fig13}.
Here only transition 10 remains at $B > 0.5~\textmd{T}$.
Thus, both models give the same result: only transitions $1-3$ and $10$
remain at a strong magnetic field. However, the HPB model is advantageous,
being simple and easy for calculations.

\section{Conclusion}
It is demonstrated that simple and efficient ``$\lambda$-method" and ``$\lambda/2$-method" based
on nanometric-thickness cells filled with alkali metal atoms allow to study
behavior of atomic Zeeman transitions of $^{85}$Rb, $^{87}$Rb~$D_1$ lines in a wide range
of magnetic field from $1$~mT to $1$~T. Particularly, for the case of  $\sigma^{+}$ polarized
laser radiation and $B > 0.5~\textmd{T}$, only 6 transitions remain in the transmission
spectrum of $^{85}$Rb~$D_1$ line, and  only $4$ transitions remain in $^{87}$Rb spectrum.
For $B > 0.6$~T the expression, which is valid in the frame of uncoupled basis
(hyperfine Paschen-Back regime), describes very well the experimental results
 for $^{85}$Rb atomic transitions. The latter is important for the determination of
 such parameters as: the atomic transitions frequency position and frequency
 separation of the components; the slope $S$ in dependence of atomic transition
 components frequency on magnetic field can be easily calculated with an inaccuracy
 of $2\%$. For $^{87}$Rb having larger hyperfine splitting, the experimental results are
 very well described in the frame of coupled basis, meanwhile the uncoupled basis
  model yields inaccuracy of $\sim$ 10$\%$ for the range of $0.5 - 0.7~\textmd{T}$. Consistency of
  the two models for $^{87}$Rb are expected to reach at $B \geq 1$~T.\\
\indent It is worth noting that calculations of magnetic field dependence of
Zeeman transition probabilities and frequency positions for the case
of  $\sigma^{+}$ polarized laser radiation performed in the frame of
the coupled basis model are fully consistent with experimental results for all the
atomic transitions of $^{85}$Rb~$D_1$ line (twenty transitions) and $^{87}$Rb~$D_1$
line (twelve transitions) in a broad range of magnetic field ($1~\textmd{mT} - 1~\textmd{T}$).
Such calculations will be of interest also for Cs, K, Na, Li.\\
\indent The results of this study can be used to develop hardware and software
solutions for magnetometers with nanometric (400~nm) local spatial resolution
and widely tunable frequency reference system based on a NTC and strong permanent magnets.\\

\begin{acknowledgments}
The research leading to these results has received funding from the European
Union FP7/2007-2013 under grant agreement no. 295025-IPERA. Research in part, conducted in the
scope of the International Associated Laboratory  (CNRS-France $\&$ SCS-Armenia)~IRMAS. Authors
A. S., G. H, and D. S thank for the support by State Committee
Science MES RA, in frame of the research project N\textsuperscript{o}. SCS 13-1C029.
\end{acknowledgments}

\end{document}